\documentclass[%
 reprint,
 superscriptaddress,
nofootinbib,
 amsmath,amssymb,
 aip,
]{revtex4-2}

\usepackage{graphicx}% Include figure files
\usepackage{dcolumn}% Align table columns on decimal point
\usepackage{bm}% bold math
\usepackage{hyperref}% add hypertext capabilities
\usepackage{physics}
\usepackage{float}
\usepackage{xcolor}

\usepackage{lipsum}

\begin{document}

\preprint{APS/123-QED}

\title{Deep Learning the Small-Angle Scattering of Polydisperse Hard Rods}% Force line breaks with \\
\author{Lijie Ding}
\affiliation{Neutron Scattering Division, Oak Ridge National Laboratory, Oak Ridge, TN 37831, USA}

\author{Changwoo Do}
\email{doc1@ornl.gov}
\affiliation{Neutron Scattering Division, Oak Ridge National Laboratory, Oak Ridge, TN 37831, USA}

\date{\today}% It is always \today, today,
             %  but any date may be explicitly specified

\begin{abstract}
We present a deep learning framework for modeling and analyzing the small-angle scattering data of polydisperse hard-rod systems, a widely used models for anisotropic colloidal particles. We use a variational autoencoder-based neural network to learn the mapping from the system parameters such as the volume fraction, rod length, and polydispersity, to the scattering function. The dataset for training and testing such neural network model is obtained from Markov chain Monte Carlo simulation of 20,000 hard spherocylinders using the hard particle Monte Carlo package from the HOOMD-blue. Four datasets were generated, each with 5,500 pairs of system parameters and corresponding scattering functions. We use one of the dataset to investigate the feasibility of the learning, and three additional datasets with different polydisperse distribution to demonstrate the generality of our approach. The neural network model transcends the fundamental limitations of the  Percus-Yevick approximation by accurately capturing anisotropic interactions and high-concentration effects that analytical models often fail to resolve. This framework achieves significantly higher accuracy in reproducing scattering functions and enables a least-square fitting routine for quantitative data analysis.
\end{abstract}

%\keywords{Suggested keywords}%Use showkeys class option if keyword
                              %display desired
\maketitle

%\tableofcontents

\section{Introduction}
% background of colloids, rod-shape particles, and SAS as a mean for characterization
Colloidal particles with anisotropic shapes,\cite{sacanna2011shape,witten2020review} such as rods, are central to soft matter physics,\cite{solomon2010microstructural} materials science, and biophysics, enabling liquid crystals,\cite{barclay1993liquid} gels,\cite{tohyama1981network} and self-assembled structures\cite{stupp1998self,liu2011self} with tailored optical, mechanical, and rheological properties. Rod-like colloids serve as versatile model systems for understanding packing behavior, excluded-volume interactions, and collective dynamics in anisotropic environments, while their polydispersity in length and diameter closely mirrors real-world materials, including filamentous viruses,\cite{dogic2006ordered} carbon nanotubes,\cite{popov2004carbon} gold nanorods,\cite{perez2005gold} and other synthetic nanorods\cite{dai1995synthesis,aspoukeh2022synthesis} widely used in plasmonics,\cite{zheng2021gold} biomedicines,\cite{yang2015gold} and catalysis.\cite{paxton2004catalytic}

% challenges on SAS for rod, lack of analytical solution
Small-angle scattering (SAS),\cite{lindner2002neutrons,chen1986small} such as small-angle X-ray scattering (SAXS)\cite{li2016small,chu2001small} and small-angle neutron scattering (SANS)\cite{shibayama2011small,windsor1988introduction} are indispensable tools for characterizing the structure of these colloidal systems at the nanoscale. These techniques deliver information on particle dimensions, size distributions, and interparticle interactions across length scales from a few nanometers to hundreds of nanometers. Understanding the structural features through SAS is essential for designing and optimizing colloidal materials. While spherical models are well-established, many high-performance materials rely on anisotropic building blocks to achieve specific mechanical or optical properties.  However, the transition from simple spheres to these rod-like colloidal systems introduces significant theoretical challenges. Specifically, there is a lack of exact analytical solutions for the scattering functions of polydisperse rods, which hinders the quantitative analysis of experimental SAS data. Conventional approximation models, such as the Percus-Yevick approach, often fail to capture the complex anisotropic interactions between rods and tend to deviate significantly as the system reaches higher concentrations.

% machine learning for SAS analysis: simulation + ML 
Recent works in machine learning (ML)-assisted SAS data analysis\cite{chang2022machine,ding2025colloids,ding2025hardsphere,tung2025scatterin_lamellar,tung2025insights_lamellar,tung2024inferring_lamellar,ding2024mechanical_polymer,ding2025charge_polymer,ding2025ladder_polymer,ding2025deciphering_polymer} offer a data-driven way to address these challenge. By leveraging extensive simulation datasets, ML models can learn complex, non-linear mappings between system parameters (such as volume fraction, particle dimensions, and polydispersity) and the corresponding scattering functions, even in regimes where analytical approximations break down. This approach has been applied to a wide range of soft matter systems including colloids,\cite{chang2022machine,ding2025colloids,ding2025hardsphere} lamellar,\cite{tung2025scatterin_lamellar,tung2025insights_lamellar,tung2024inferring_lamellar} and polymers.\cite{ding2024mechanical_polymer,ding2025charge_polymer,ding2025ladder_polymer,ding2025deciphering_polymer} It enables accurate forward prediction, parameters inversion, and robust interpretation of scattering profiles without relying on conventional approximation methods.

% what we do in this work
In this work, we extend this simulation-informed ML framework to polydisperse hard rods system. By training a variational autoencoder\cite{doersch2016tutorial} (VAE)-based neural network using synthetic data generated by Markov chain Monte Carlo\cite{krauth2006statistical} (MCMC) simulations for hard spherocylinders, we establish an accurate and robust mapping between the system parameters and the scattering function. We first use principal component analysis to study the feasibility of learning the mapping for each system parameter, such as volume fraction $\phi$, mean rod length $L$, length polydispersity $\sigma_L$, and diameter polydispersity $\sigma_D$. We then train the neural network model to obtain a generator model to directly generate the scattering function $I(Q)$ from the system parameters $(\phi,L,\sigma_D)$ for polydisperse distributions including uniform, normal, and lognormal. We show that the neural network model achieve much lower error comparing with the conventional approximation model. Finally, we use the generator model to fit scattering function using least-square regression and show that the fitting routine is robust and the fitted parameters are accurate.

% structure of this paper
The rest of this paper is organized as follows. In Sec.~\ref{sec:method}, we introduce the MCMC simulation we use, the scattering analysis we implement, and the design and training of our neural network architecture. We present the results for the data analysis, comparison of generation function, and fitting routine in Sec.~\ref{sec:summary}. Finally, we summarize this work in Sec.~\ref{sec:summary}.

\section{Method}
\label{sec:method}

\subsection{Monte Carlo Simulation}

%1 overview, hard rods MC simulation, hoomd-blue
To calculate the scattering of the polydisperse hard rods system, we carry out MCMC\cite{krauth2006statistical} simulations using HOOMD-blue.\cite{anderson2020hoomd, anderson2016scalable} The hard rods are modeled by the hard spherocylinders, and the simulation is carried out in the canonical (NVT) ensemble with periodic boundary condition.

%2 more detail, monte carlo move, how polydispersity is modeled, initialization routine.
We employed the hard particle Monte Carlo (HPMC) simulation package\cite{anderson2020hoomd} within HOOMD-blue to represent the hard spherocylinders via the convex spheropolygon integrator. In this framework, the cylindrical axis is defined using two vertices separated by the rod length $L_i$, combined with a finite sweep radius corresponding to half of the particle $D_i$. The simulation incorporates independent polydispersity for both rod length and diameter, accounting for three distinct probability distributions: uniform, normal, and lognormal. The HPMC's convex spheropolygon integrator handles the particle interactions as hard-core, pure repulsive, such that there is a zero potential for non-overlapping configurations and infinite overlap penalty. The configuration updates include two kinds of trial moves: a random rigid-body translation with a automatically tuned maximum displacement, as well as a random rotation about the particle center also with automatically tuned maximum rotation angle. The HOOMD-blue's built-in move size optimization is used to target acceptance probabilities of 0.3 for these updates. The system is initialized by 
placing all rods on a simple cubic lattice in the simulation box of size matching initial volume fraction $\phi_{init}=0.01$, with all rods pointing along the z direction. The system is then randomized and compressed into the target volume fraction through HPMC's quick compression with a quadratic progression such that the system compresses faster at early stage and slower later. After the compression, the simulation is randomized and then moved on to do MCMC sampling for the scattering function measurement $I(Q)$.

%3 some parameters, system size, simulation steps, etc
Systems of $N=20,000$ polydisperse hard rods are simulated. Randomization before and after the compression stage is short runs of 2,000 time steps each. The sampling for $I(Q)$ measurement are from 100 independent configurations, taken at intervals of 1,000 time steps afterwards. Each time step consists of 4 trial moves per particle, on average 2 translations and 2 rotations. We generate four datasets using the MC simulation, each consists of 5,500 independent simulation runs: one with uniform distributed rod length and diameter for the polydispersity $\sigma_L,\sigma_D\in[0,0.3]$ uniformly among all of the run, while the remaining three datasets explore different type of polydispersity distribution (uniform, normal, and lognormal) but with $\sigma_L=0$. The volume fraction $\phi\in[0.01,0.3]$ and mean length $L\in[0.5,5]$ are uniformly distributed among all of the simulation run. In practice, we use 70\%-30\% split of the data for training and testing.

\subsection{Small-angle scattering analysis}
For anisotropic particles like the rods, the structure factor $S(Q)$ is coupled with the particle form factor $P(Q)$, thus we calculate the normalized scattering function of the polydisperse hard rods directly using\cite{lindner2002neutrons,chen1986small}:

\begin{equation}
    I(\vb{Q}) = \left< \frac{A(\vb{Q})A^*(\vb{Q})}{\sum_i V_i^2}\right> 
    \label{eq:IQ}
\end{equation}
where $\vb{Q}$ is the scattering vector, $V_i=\pi D_i^2 L_i/4 + \pi D_i^3/6$ is the volume of rod $i$, summation $\sum_i$ is over all rods, $\left<\dots\right>$ is the ensemble average of all configurations, $A(\vb{Q})$ is the scattering amplitude given by:
\begin{equation}
    A(\vb{Q}) = \sum_i e^{i\vb{Q}\vb{r}_i} F(Q;\alpha,L_i,D_i)
    \label{eq:AQ}
\end{equation}
where $\vb{r}_i$ is the position of particle $i$, $F(Q;\alpha,L-i,D_i)$ is the scattering amplitude of the hard rods, given by the capped cylinder model with cap radius equals to the cylinder radius \cite{kaya2004scattering, kaya2004scattering_add}:
\begin{equation}
    \begin{aligned}
    &F(Q,\alpha,L_i,D_i) = \frac{\pi D_i^{2}L_i}{4} \mathrm{sinc}\left(w \right) \frac{2J_{1}\bigl(v\bigr)}{v} + \\
    &\pi D_i^{3}\int_{0}^{1}(1-t^{2})\frac{J_{1}\bigl(v\sqrt{1-t^{2}}\bigr)}{v\sqrt{1-t^{2}}}
   \cos\left[w \left(\frac{D_i t}{L_i} + 1\right)\right]\mathrm{d}t  
    \end{aligned}
    \label{eq:FQ}
\end{equation}
where $D_i$ and $L_i$ are the diameter and length of the cylindrical part, $\alpha$ is the angle between the scattering $\vb{Q}$ vector and the long axis of the rod, $J_1(x)$ is the Bessel function of the first kind, $\mathrm{sinc}=\sin(x)/x$, $v = \frac{QD_i}{2} \sin\alpha$ and  $w=\frac{QL_i}{2}\cos\alpha$, . 

\begin{figure}[!t]
    \centering
    \includegraphics[width=\linewidth]{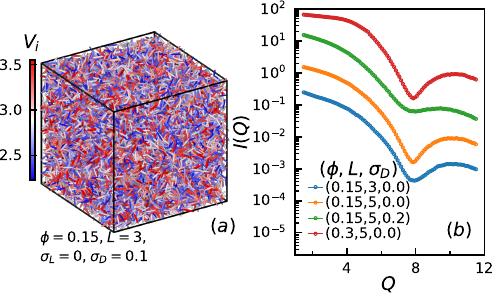}
    \caption{Illustration of scattering function of polydisperse hard rods. (a) Snapshot of the system configuration with volume fraction $\phi=0.15$, mean rod length $L=3$, length polydispersity $\sigma_L=0$, and diameter polydispersity $\sigma_D=0.1$. Color bar is the volume of each rod. (b) Scattering function $I(Q)$ with different $\phi$, $L$, and $\sigma_D$. $I(Q)$ are shifted vertically for better visualization.}
    \label{fig:Iq_and_config}
\end{figure}

Finally, the isotropic $I(Q)$ is the orientational average of $I(\vb{Q})$ over all directions. Fig.~\ref{fig:Iq_and_config} shows few examples of the scattering function $I(Q)$ of the polydisperse hard rods system, along with a sample snapshot of the system configuration.

Conventionally, the scattering function for the polydisperse rods systems can be analyzed using the Percus-Yevick approximations\cite{percus1958analysis,wertheim1963exact,kinning1984hard,katzav2019random} for the hard spheres' structure factor, combined with the form factor of the dilute polydisperse rods.
\begin{equation}
    I_{PY}(Q) = S_{PY}(Q)P(Q)
    \label{eq:IPY}
\end{equation}
The structure factor $S_{PY}(Q)= S_{PY}(Q,\phi,R_g)$ is solution from the Ornstein–Zernike equation\cite{ornstein1914accidental}, depends on the volume fraction and the radius of gyration $R_g$. The form factor $P(Q)$ for the polydisperse rods is given by the isotropic average of the scattering amplitude of the rods with targeted size distribution.
\begin{equation}
    P(Q) = \frac{\sum_i\int_0^{\pi/2} |F(Q,\alpha,L_i,D_i)|^2 \sin\alpha \mathrm{d}\alpha}{\sum_i V_i^2}
    \label{eq:PQ}
\end{equation}

Detail of the mathematical formula of $S_{PY}(Q)$ can be found in the references\cite{percus1958analysis,wertheim1963exact,kinning1984hard}, which originally depend on the radius of the hard sphere $R_s$. To adopt it for the rods, we substitute the sphere radius with radius of gyration $R_s = \sqrt{5/3}R_g$, and use the volume-weighted $R_g$ of the polydisperse hard rods system. The traditional method misses the coupling between the form factor and the inter-particle structure factor due to both the polydispersity and the rods' orientation, and will be used for comparison with our approach.

\subsection{Variational Autoencoder-based neural network}
\begin{figure}[!t]
    \centering
    \includegraphics[width=\linewidth]{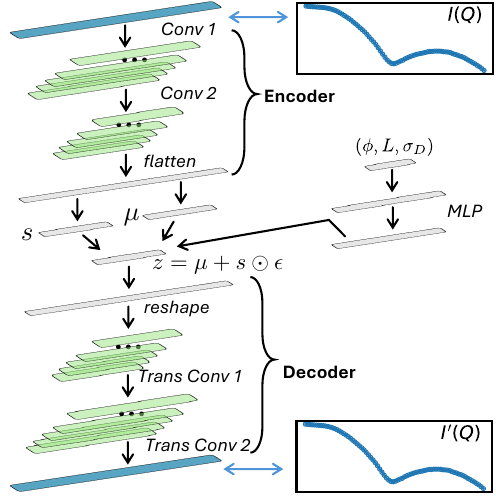}
    \caption{Overview of the neural network architecture for learning the mapping between the system parameters $(\phi, L,\sigma_D)$ and the scattering function $I(Q)$. The neural network consist of three parts, in which an encoder with 2 convolutional layers and the symmetric decoder compress the input scattering function into a low dimensional latent space, a multilayer perceptron learns the mapping between the system parameters and the latent variables.}
    \label{fig:NN_architecture}
\end{figure}
Despite the lack of closed-form expression for the polydisperse hard rods system, it is possible to learn the mapping between system parameters and the scattering function using simulation data. To achieve this, we use a variational autoencoder based neural network, as shown in Fig.~\ref{fig:NN_architecture}. The neural network consists of three parts: an encoder with 2 convolutional layer\cite{o2015introduction} that transform the high-dimensional input scattering function into a low-dimensional latent space, a decoder that decodes the latent variables back to the scattering function, and a multilayer perceptron (MLP) that map the system parameters and the latent variables. 

% more on the three parts: encoder, decoder, and mlp
The encoder and decoder together make up a VAE\cite{doersch2016tutorial} for compressing the scattering function into the latent space, which is trained on the loss function:
\begin{equation}
    L_{VAE} =\frac{1}{N} \sum_{I(Q)} \left<\left[\log_{10}I(Q) - \log_{10}I'(Q) \right]^2 \right>_{Q}
    \label{equ:loss_VAE}
\end{equation}
that measure the Euclidean distance between the log of the input scattering function $I(Q)$ and the output $I'(Q)$, average over all $Q$ value through $\left<\dots\right>_Q $, and $N$ samples. The input $I(Q)$ is transform to the latent mean variable $\mu$ and standard deviation variable $s$, together reparameterized into $z=\mu + s\odot \epsilon$ using the normally distributed variable $\epsilon\sim \mathcal{N}(0,s)$, sampled from normal distribution. The reparameterized $z$ is transformed to the output $I'(Q;\epsilon)$, and we get the reconstructed scattering function $I(Q)=\left<I'(Q;\epsilon)\right>_\epsilon$ by averaging with 100 randomly sampled $\epsilon$.

The MLP transform the input system parameters to the latent variable, which can then be transformed to the output scattering function $I'(Q)$ using the decoder. The MLP and the decoder together constitute a generator that generates the scattering function $I'(Q)$ directly from the system parameters. The generator is trained with the same loss function as in Eq.\eqref{equ:loss_VAE}

% more detail on parameter of the neural network, and the training procedure.
The two convolutional layers in the encoder are of kernel size $9$ and stride 2. The first layer has 30 channels, and the second layer has 60. These two convolutional layers transform the 100 dimensional input $I(Q)$ into a $60\times 25=1500$ dimensional vector, which is then transformed to the 3 dimensional latent variables. The decoder is made of two transposed convolutional layers symmetric to the ones in the encoder. The MLP consists of two linear layer of size 9, with ReLU as activation function. To train this neural network, we firstly train the VAE alone for 2000 epoch, then train the MLP while freezing the decoder for 300 epoch, finally the generator including the MLP and the decoder is fine tuned for another 300 epoch. In practice, we implement the neural network using PyTorch\cite{paszke2019pytorch} and train it using Adam optimizer\cite{kingma2014adam} with CosineAnnealingLR scheduler\cite{loshchilov2016sgdr}.

\section{Results}
\label{sec:results}
\subsection{Scattering function of hard rods}
We firstly study the scattering function $I(Q)$ of the hard rods system, and investigate the feasibility for the generation function to learn the mapping between the system parameters and the scattering function by carrying out principal component analysis of the dataset.

\begin{figure}[!t]
    \centering
    \includegraphics[width=\linewidth]{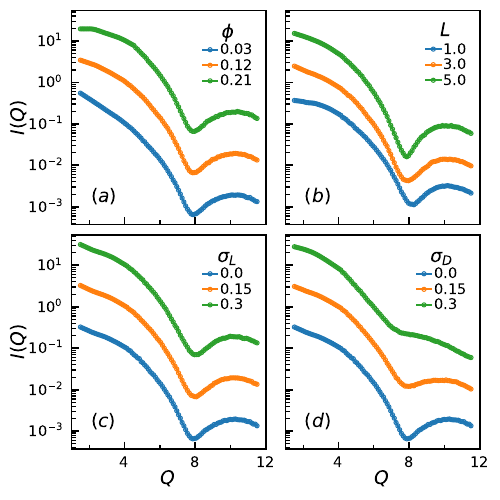}
    \caption{Examples of the scattering function $I(Q)$ for polydisperse hard rods with uniform distributed length and diameter size $L_i\in U(1-\sigma_L,1+\sigma_L), D_i\in U(1-\sigma_D,1+\sigma_D)$, for various volume fraction $\phi$, mean length $L$ and polydispersity $\sigma_L,\sigma_D$. Default values are $(\phi,L,\sigma_L,\sigma_D)=(0.15,2,0,0)$. (a) Scattering function $I(Q)$ for different volume fraction $\phi$ with other parameters fixed. (b) $I(Q)$ for different mean length $L$. (c) $I(Q)$ for different length polydispersity $\sigma_L$. (d) $I(Q)$ for different diameter polydispersity $\sigma_D$.}
    \label{fig:Iq_example}
\end{figure}

Fig.~\ref{fig:Iq_example} shows the variation of the scattering function $I(Q)$ respect to each system parameters, individually. The volume fraction affect more on the low $Q$ region where the rod-rod interaction are captured, the change of mean length $L$ is reflected on the slope of $I(Q)$ in the mid $Q$ range, and the diameter polydispersity $\sigma_D$ influences mostly the dip of the $I(Q)$ near the $Q$ corresponding to the diameter, whereas the change of length polydispersity $\sigma_L$ is barely captured by the scattering function.

\begin{figure}[!t]
    \centering
    \includegraphics[width=\linewidth]{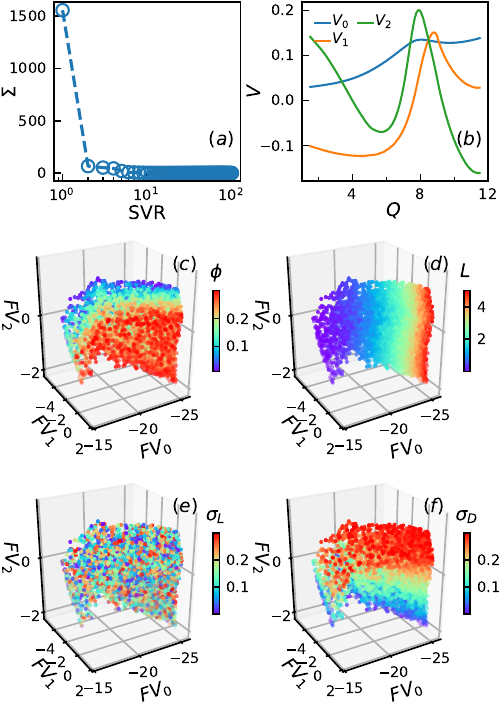}
    \caption{Principal component analysis of the scattering function dataset for $\vb{F} = \{\log_{10}{I(Q)}\}$ of uniform-distribution polydisperse hard spheres. (a) Decay of the singular value entry in $\Sigma$ in $\vb{F}=\vb{U}\vb{\Sigma}\vb{V}^T$ versus the singular value rank (SVR). (b) First 3 singular vectors $(V0,V1,V2)\in\vb{V}$ . (c) Distribution of the volume fraction $\phi$ in the singular value space $(FV0,FV1,FV2)$ in which each $\log_{10}(I(Q))\in \vb{F}$ is projected to the $(V0,V1,V2)$. (d)-(f) Distribution of the mean rod length $L$, length polydispersity $\sigma_L$, and diameter polydispersity $\sigma_D$, respectively.}
    \label{fig:svd_combined}
\end{figure}

To further investigate the feasibility of learning the mapping between the system parameters $(\phi,L,\sigma_L,\sigma_D)$ and the scattering function $I(Q)$, we carry out principal component analysis of the dataset $\vb{F}_0=\{\log_{10}{I(Q)}\}$ containing variation of all four system parameters. The singular values in Fig.~\ref{fig:svd_combined}(a) shows the quick decay respect to the singular value rank (SVR), which indicate that most information about the $I(Q)$ can be represented by the first few singular vectors, of which the first three are shown in Fig.~\ref{fig:svd_combined}(b). By projecting the all of the $I(Q)$ in the dataset $\vb{F}$ on to the first three singular vectors $(V_0,V_1,V_2)$, the $I(Q)\in\vb{F}$ can be represented by a vector in $\mathcal{R}^3$. Fig.~\ref{fig:svd_combined}(c)-(f) show the distribution of all four system parameters in this projected space. Among these, the distribution for the volume fractions $\phi$, mean rod length $L$, and diameter polydispersity $\sigma_D$ are ordered, indicating the feasibility for mapping these parameters and the scattering function. On the contrary, the distribution of the length polydispersity $\sigma_L$ in the singular vector space is random, making it unsuitable for the mapping in the current setting. Thus moving on, we will carry out the rest of the work using the dataset $\vb{F}$ containing various $\phi$, $L$, and $\sigma_D$ only.

\begin{figure}
    \centering
    \includegraphics[width=\linewidth]{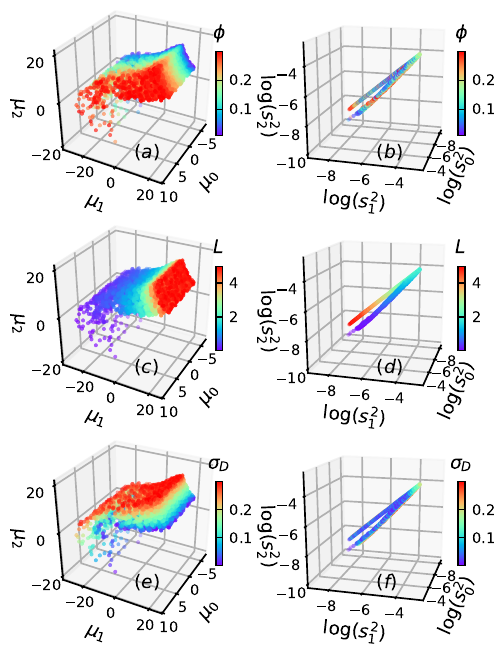}
    \caption{Distribution of latent variables $\mu$ and $\log(s^2)$ for the scattering function of rods with uniformly distributed diameter. (a), (c) and (e) are for the distribution of volume fraction $\phi$, mean length $L$, and diameter polydispersity $\sigma_D$ in the $\mu$ space, respectively. (b), (d), (f) are for the corresponding system parameters in the $\log(s^2)$ space.}
    \label{fig:latent_space_distribution}
\end{figure}

By training the VAE part of the neural network, we can obtain the latent representation of the scattering function. Similar to the representation of the system parameters in the singular vector space, we can also visualize the distribution of these system parameters $(\phi,L,\sigma_D)$ in the latent $\mu$ and $s$ space, respectively. Fig.~\ref{fig:latent_space_distribution} shows these distributions. The distribution of the three system parameters within the latent  mean $\mu$ space exhibits a clear, organized topology, suggesting that the VAE has successfully encoded the underlying physics of the scattering profiles. Furthermore, the latent log-variance $\log(s^2)$ reaches values as low as approximately $-6$, indicating most of the information are stored in the latent mean space.

\subsection{Generation of scattering function}
We then study the generator of neural network that combine the MLP and the decoder. The generator directly generate the scattering function $I(Q)$ from the system parameters $(\phi,L,\sigma_D)$ with one forward pass in the neural network, bypassing the need to hours-long MC simulations. We compare the accuracy of the neural network-based generator model and the conventional Percus-Yevick approximation respect to the simulation-calculated $I(Q)$.

\begin{figure}[!t]
    \centering
    \includegraphics[width=\linewidth]{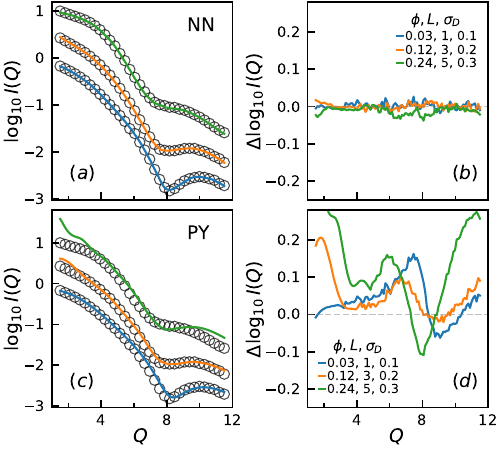}
    \caption{Comparison between the scattering function $I(Q)$ calculated from the simulation and the corresponding ones $I'(Q)$ from the neural network (NN) model and the Percus-Yevick (PY) approximation. (a) Simulation calculated $I(Q)$ versus the neural network generated $I'(Q)$. (b) Difference $\Delta\log_{10}{I(Q)}=\log_{10}{I'(Q)/I(Q)}$ between the simulation $I(Q)$ versus the neural network generated $I'(Q)$. (c) and (d) are similar to (a) and (b) but for the Percus-Yevick approximation.}
    \label{fig:gen_vs_PY_Iq_uni}
\end{figure}

Fig.~\ref{fig:gen_vs_PY_Iq_uni} shows the examples of the comparison between the generated $I'(Q)$ versus the simulation calculated $I(Q)$, for both the neural network model, and the Percus-Yevick approximation. As shown in Fig.~\ref{fig:gen_vs_PY_Iq_uni}(a) and (b), the neural network generated scattering function matches the simulation results closely, with very small difference quantified by $\Delta\log_{10}{I(Q)}=\log_{10}{I'(Q)/I(Q)}$. On the contrary, the Percus-Yevick approximation deviate from the reference $I(Q)$ near the minimum at low volume fraction and the deviation amplifies rapidly with increasing volume fraction, mean rod length, and polydispersity.

\begin{figure}[!t]
    \centering
    \includegraphics[width=\linewidth]{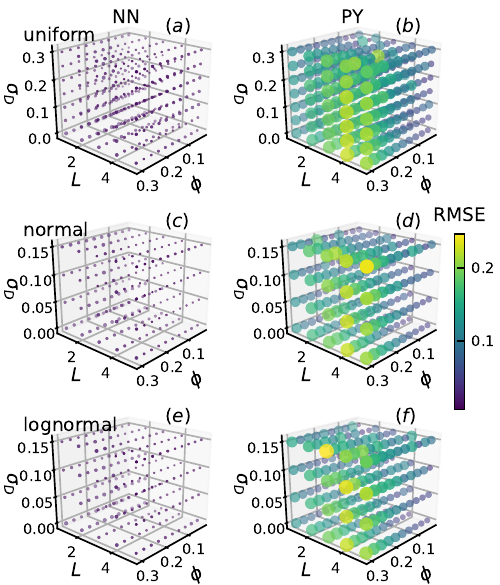}
    \caption{Root mean square error (RMSE) of the generated $I'(Q)$ using both neural network (NN) model and Percus-Yevick (PY) approximation for different system parameters, for three kinds of polydispersity distributions. (a) RMSE of the NN model for different volume fraction $\phi$, mean rod length $L$, and diameter polydispersity $\sigma_D$ with uniform distribution. (b) RMSE of the PY approximation for different $\phi$, $L$, and $\sigma_D$ with uniform distribution. (c) and (d) similar to (a) and (b) but for polydispersity with normal distribution. (e) and (f) similar to (a) and (b) but for polydispersity with lognormal distribution.}
    \label{fig:gen_vs_PY_MSE_uni}
\end{figure}

Moreover, Fig.~\ref{fig:gen_vs_PY_MSE_uni} shows the Root mean square error $RMSE=\sqrt{\left< (\log_{10}{I'(Q)/I(Q)})^2\right>_Q}$ for both methods cross various system parameters $(\phi,L,\sigma_D)$, for all three kinds of polydispersity distribution including uniform, normal, and lognormal. Overall the neural network generator stay at low error across the parameters regions, whereas the Percus-Yevick approximation introduce relatively large error, especially at high volume fraction, long rod length, and high polydispersity. The accuracy of the neural network model, combined with its speed due to single forward pass, make it suitable for fitting the scattering data using a least square fitting routine.

\subsection{Fitting the scattering}
Leveraging the neural network generator, we can carry out least-square fitting routines for the scattering data which was previously only possible for systems with analytical solutions. We demonstrate the robustness of such fitting routine by running the fitting algorithm with different initial values, and test the accuracy by fitting all of the scattering function in the test dataset for all three kinds of polydisperse distribution.

\begin{figure}[!t]
    \centering
    \includegraphics[width=\linewidth]{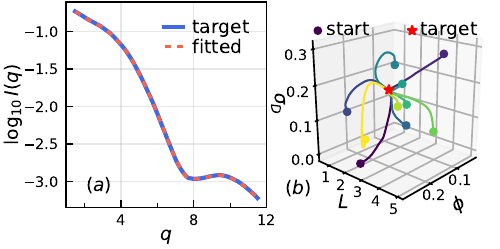}
    \caption{Fitting the target scattering function $I(Q)$ using the neural network generation model. (a) Comparison between the target $I(Q)$ and the fitted $I'(Q)$. (b) Convergence of the fitting parameters to the target from different initial starting points.}
    \label{fig:ls_fitting_demo}
\end{figure}

Fig.~\ref{fig:ls_fitting_demo} shows a comparison between the target scattering function $I(Q)$, and the one obtained from the least-square fitting using the neural network generator. In Fig.~\ref{fig:ls_fitting_demo}(a), it is shown that the $I'(Q)$ generated from the fitted parameters $(\phi,L,\sigma_D)$ match the target curve $I(Q)$ closely. Moreover, as shown in Fig.~\ref{fig:ls_fitting_demo}(b), the fitting procedure with different start points all converge to the same targets, illustrate the robustness of the fitting method.

\begin{figure}[!t]
    \centering
    \includegraphics[width=\linewidth]{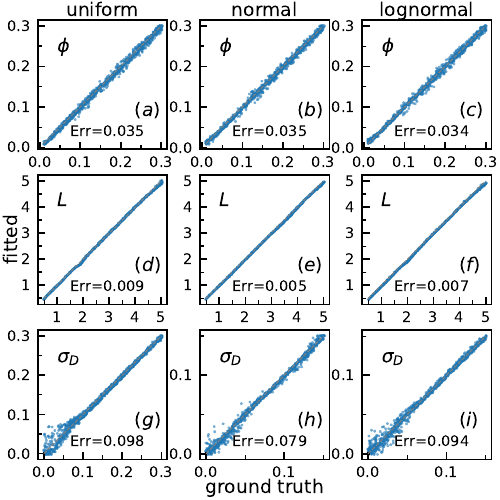}
    \caption{Fitting the test set for volume fraction $\phi$, mean rod length $L$, and diameter polydispersity $\sigma_D$ from the scattering function $I(Q)$ using the neural network (NN) model and least square fitting routine. (a),(b), and (c) are ground truth $\phi$ versus least-square fitted $\phi$ for uniform, normal, and lognormal distribution, respectively. (d)-(f) Similar to (a)-(c) but for $L$. (g)-(i) for $\sigma_D$.}
    \label{fig:LS_fitting_3_distributions}
\end{figure}

Finally, we apply the least-square fitting with the generator model that trained from the training dateset on the testing data, for all three kinds of polydispersity distribution. Fig.~\ref{fig:LS_fitting_3_distributions} shows the fitting results, comparing the ground truth for each system parameters that are used for generating the simulation calculated scattering function $I(Q)$ and the fitted ones that are obtained by fitting the $I(Q)$ using the neural network generator. The fitting routine achieved high precision cross all parameters and all polydisperse distributions, validating the effectiveness and generality of our framework.

\section{Summary}
\label{sec:summary}
% what we've done in this work
In this work, we have developed a deep learning framework to analyze the SAS of polydisperse hard rods, a fundamental model for anisotropic colloidal systems. Using MCMC simulation with HOOMD-blue, we generated comprehensive datasets of scattering functions $I(Q)$, spanning a wide parameter space  including volume fraction $\phi$, mean rods length $L$, and diameter polydispersity $\sigma_D$ for uniform, normal, and lognormal distributions. We use principal component analysis to reveal the compressibility and feasibility for mapping of the dataset, and highlighted the challenges in mapping the length polydispersity within the current study. We train a VAE-based neural network with convolutional layers and a MLP to create a generator that predicts $I(Q)$ directly from system parameters $(\phi,L,\sigma_D)$ with superior accuracy over the conventional Percus-Yevick approximation approach, especially at higher densities and polydispersity.

% what is the significance. and future directions
Our framework offers a generalizable way to analyze and interpret the SAS data for colloids systems characterized by anisotropic shape and polydispersity. Notably, this method maintains its accuracy in high concentration regimes where traditional Ornstein–Zernike approach\cite{ornstein1914accidental} lacks exact analytical solutions and common approximation methods typically break down. Furthermore, our approach can be extended to more complicated systems such as colloidal dispersion of mixed size and shapes, as well as anisotropic colloidal particles with electrostatic interactions.

\section*{Data Availability}
The code, data and trained model for this work are available at the GitHub repository \url{https://github.com/ljding94/Polydisperse_Rods}

\begin{acknowledgments}
We thank Chi-Huan Tung and Wei-Ren Chen for fruitful discussion. This research was sponsored by the Laboratory Directed Research and Development Program of Oak Ridge National Laboratory, managed by UT-Battelle, LLC, for the US DOE. This research used resources at the Spallation Neutron Source, a DOE Office of Science User Facility operated by the Oak Ridge National Laboratory. Computations used resources of the Oak Ridge Leadership Computing Facility, which is supported by the DOE Office of Science under contract No. DE-AC05-00OR22725. 
\end{acknowledgments}

% The \nocite command causes all entries in a bibliography to be printed out
% whether or not they are actually referenced in the text. This is appropriate
% for the sample file to show the different styles of references, but authors
% most likely will not want to use it.
%\nocite{*}

\section*{Reference}
\vspace{-15pt}
\bibliography{reference}% Produces the bibliography via BibTeX.

\end{document}